# Hidden Phases Revealed at the Surface of Double-Layered $Sr_3(Ru_{1-x}Mn_x)_2O_7$

Chen Chen,[1] Jisun Kim,[1] V. B. Nascimento,[2] Zhenyu Diao,[1] Jing Teng,[1] Biao Hu,[1] Guorong Li,[1] Fangyang Liu,[1] Jiandi Zhang,[1] Rongying Jin,[1] and E. W. Plummer[1]*

[1]*Department of Physics & Astronomy, Louisiana State University, Baton Rouge, Louisiana 70803, USA*

[2]*Departamento de Física, ICEx-UFMG, CP 702, Belo Horizonte, MG CEP 30123-970, Brazil*

# Abstract

Double-layered $Sr_3Ru_2O_7$ has received phenomenal consideration because it exhibits a plethora of exotic phases when perturbed. New phases emerge with the application of pressure, magnetic field, or doping. Here we show that creating a surface is an alternative and effective way to reveal hidden phases that are different from those seen in the bulk by investigating the surface properties of $Sr_3(Ru_{1-x}Mn_x)_2O_7$. Driven by the tilt distortion of $RuO_6$ octahedra, the surface of $Sr_3Ru_2O_7$ is less metallic than the bulk. In contrast, because of the vanishing of tilt and enhanced rotation with Mn-doping, the surface of $Sr_3(Ru_{0.84}Mn_{0.16})_2O_7$ is metallic while the bulk is insulating. Our result demonstrates that the electronic and structural properties at the surface are intimately coupled and consistent with quasi two-dimensional character.

The Ruddlesden-Popper (RP) series $Sr_{n+1}Ru_nO_{3n+1}$ has attracted considerable attention because of its diverse and complex phases, where the ground state can be tuned by changing the number of $RuO_6$ octahedral layers ($n$) in the unit cell, chemical doping, or external parameters such as magnetic field and pressure [1-11]**Error! Bookmark not defined.Error! Bookmark not defined.**. For example $Ca_{2-x}Sr_xRuO_4$ ($n$=1) is a $p$-wave superconductor when $x$=2 [5,12], and an antiferromagnetic (AFM) insulator when $x$=0 [13,14]. When uniaxial pressure is applied to $Sr_2RuO_4$ the superconducting transition temperature increases [15]. These physical properties are closely coupled with the structure: rotation and tilt of the $RuO_6$ octahedron [16-22]. The double-layered paramagnetic and metallic $Sr_3Ru_2O_7$ ($n$=2), shown in Fig. 1(a), also exhibits physical properties sensitive to perturbation. $Sr_3Ru_2O_7$ shows metamagnetic quantum criticality under magnetic field [2,4,6], and pressure-induced ferromagnetic (FM) order [23,24]. Doping $Sr_3Ru_2O_7$ with Ca or Mn induces new phases [25,26]. With Ca doping the system goes from itinerant metamagnetic to the AFM phase, accompanied by a transition from metallic to Anderson localized state. These transitions are associated with the increase of octahedral rotation and onset of tilt [26-29]. Partial replacement of Ru by Mn leads to a metal-insulator transition (MIT) whose transition temperature $T_{MIT}$ increases with increasing Mn concentration [25]. This system undergoes a transition from a paramagnetic metal at high temperatures to a long range ordered antiferromagnetic phase at low temperatures for $x$~0.16 [30]. In the bulk, both the electronic and magnetic properties are closely coupled to the evolution of the structural distortion (octahedral rotation).

Creating a surface is an effective perturbation to tip the subtle balance of competing interactions between the structural, charge, orbital, and spin degrees of freedom, thus offering an opportunity for revealing phases hidden in the bulk. For instance, the previous studies of $Ca_{2-x}Sr_xRuO_4$ showed that cleaving this material enhances $RuO_6$ rotational distortion and induces a buckling of the Ca/Sr-O surface plane, which suppresses electronic phase transition temperatures but avoid simultaneous lattice distortion, thus producing a purely electronic Mott transition for $x$=0.1 [31], as well as a much different phase diagram than the bulk [32]. It is understood from the previous work that $RuO_6$ rotation is enhanced and tilt is induced at the surface of $Sr_3(Ru_{1-x}Mn_x)_2O_7$ [33], which leads to a different electronic behavior [34]. Here we establish the surface structure-property relationship.

The bulk tetragonal (tet) unit cell for the bilayer $Sr_3(Ru_{1-x}Mn_x)_2O_7$ is shown in Fig. 1(a). There is bilayer octahedra stacking along the *c*-axis direction, with the Ru/Mn atoms located at the center of the octahedron. For the parent compound there is octahedral rotation, shown in a planar view in Fig. 1(b). In the bulk the in-plane $RuO_6$ rotation in the second layer is opposite to the first layer, indicated by the light grey octahedra in Fig. 1(b) [23,35]. The rotation creates a new unit cell, shown by the black square in Fig. 1(b), denoted as $\left(\sqrt{2}\times\sqrt{2}\right)\mathrm{R}45^{\circ}$, with respect to the tet-$(1\times1)$ unit cell indicated by the red dashed square in Fig. 1(b). The black arrows illustrate that there is left and right-handed chirality associated with the in-plane rotation of the octahedra. This chirality has been imaged using the scanning tunneling microscopy (STM) [34], as displayed in Fig. 1(c) for $x$=0.06. Samples cleave at the weak bonding between the two stacks of the double-layered octahedra, indicated by the dashed line in Fig. 1(a), thus exposing the SrO layer. The ability to see the subsurface Mn atoms is a consequence of the tilt distortion of $(Ru/Mn)O_6$ octahedra at the surface [33], which reduces the symmetry from $C_{4v}$ to $C_{2v}$, changing the electronic properties of the surface with respect to the bulk [34]. Fig. 1(d) depicts a tilt distortion in a bilayer perovskite, where the octahedra rotate out of the plane around the axis parallel to the octahedral edge, and the top and bottom octahedra are forced to tilt in opposite directions because they share one oxygen atom.

Here we show that the surface phase diagram is dramatically different from the bulk due to enhanced rotation of octahedra, coupled with a surface induced octahedral tilt which is a function of Mn doping level. The presence of tilt makes the surface less metallic while the bulk is conducting for low Mn doping. The removal of tilt with remaining rotation at the surface for higher Mn doping makes the surface more metallic while the bulk is insulating with no octahedral rotation. The structural and metallicity evolution with Mn concentration at the surface exhibits linear behavior, which is consistent with the behavior of an order parameter of a surface second order transition, in sharp contrast to the structural transition in the bulk.

In this study we utilize a combination of surface sensitive techniques in order to determine the surface phases. These include low energy electron diffraction (LEED) intensity-voltage (*I-V*) analysis to probe the symmetry and the surface octahedral distortion, and high resolution electron energy loss spectroscopy (HREELS) to explore surface metallicity by examining the electron-phonon coupling. The single crystals used here were grown by the floating-zone technique in an image furnace, with details described elsewhere [25]. The samples for each doping level were from the same growth batch for all measurements. All the samples were cleaved at 86 K in an ultra-high vacuum (UHV) environment with base pressure < $1\times10^{-10}$ Torr, and immediately transferred in-situ to the characterization chamber. Both LEED and HREELS data were collected within 30 minutes after cleaving. LEED experiments were conducted in a μ-metal shielded chamber. Diffracted patterns from elastically scattered electrons were collected within an energy range of 50-360 eV using a home-built video-LEED camera data collecting system. Both *I-V* curves and line-profile curves were extracted from digitalized diffraction patterns. The HREELS data were collected in the same vacuum chamber as the LEED using a LK-5000, with double magnetic shielding. A focused-monochromized low energy electron beam impinges upon the sample, and the scattered electrons are resolved in energy and momentum. The measurements here were taken in the specularly reflected direction ( 70° incident angle), where the 7 eV incoming electrons interact with the long range dipole field in the vacuum. This makes the HREELS measurement very surface sensitive.

Fig. 2(a) shows a schematic LEED pattern for a surface with octahedral rotation but no tilt, which is the case for both $Sr_2RuO_4$ [36] and $Sr_3(Ru_{1-x}Mn_x)_2O_7$ ($x$ > 0.16) surface. Using the tet-$(1\times1)$ as the basis (Fig. 1(a)), the $\left(\sqrt{2}\times\sqrt{2}\right)$R45° spots are labeled fractional, e.g. (3/2,3/2) spot indicated by the black dashed circle in Fig. 2(a). With the symmetry ($C_{4v}$) shown in Fig. 1(b) the LEED pattern has two glide lines (dotted in Fig. 2(a)), where the fractional ordered spots are missing (dotted circles for (3/2,3/2) and equivalent spots) [32,36]. Fig. 2(b) shows the LEED pattern taken at 86 K and 225 eV for the parent $Sr_3Ru_2O_7$, where the existence of tilt has broken the $C_{4v}$ symmetry seen in Fig. 2(a) for only rotation. Tilt breaks the glideline symmetry along one direction, thus the LEED pattern symmetry is lower ($C_{2v}$) than four-fold $C_{4v}$. The green circles show the locations of the two diffracted spots along the broken glideline (forbidden for the rotation only case), and the red circles show the locations of the two extinguished spots along the existing glideline. The purple and orange circles mark fractional ordered spots which have different intensities because of the broken symmetry.

LEED *I-V* structural analysis was performed using only the fractional order spots exploiting a procedure used previously [33]. Fractional ordered spots are more sensitive to rotation and tilt than the integer spots. Because of the low surface Debye temperature, the factional beam intensity cannot be clearly resolved beyond 360 eV [37]. The Debye temperatures used here are 550 K for O, 600 K for Ru, and 145 K for Sr. Prior to analysis, the LEED *I-V* data is smoothed using the Savitzky-Golay method with the 3$^{rd}$-order polynomial. Each *I-V* curve is a sum of at most 4 equivalent spots. For example, (1/2,3/2) is obtained from the sum of the (1/2,3/2), (1/2,-3/2), (-1/2,3/2), and (-1/2,-3/2) spots.

Agreement between the calculated and experimental *I-V* profiles as a function of the assumed structure is quantified by using the Pendry reliability factor ($R_P$) [38]. The fitting procedure adapts a simulated annealing global searching algorithm. The unit cell was relaxed following the atomic movements that maintain the glide plane symmetry of the lattice. Then the rotation and tilt angles are determined by optimizing the $R_P$ value. The experimental and optimized LEED *I-V* curves are presented in Fig. 3(a). The associated error in the structural parameters is estimated with the equation $\sigma = (R_P^{Total})_{\min}\sqrt{\frac{8|V_{oi}|}{\Delta E}}$ proposed by Pendry [38]. The total energy range $\Delta E$ is 759 eV, and $V_{oi}$ the imaginary part of the inner potential is ~ 6 eV. With the total $(R_P^{Total})_{\min}$ factor of 0.28, the estimated error is $\sigma = 0.07$. Fig. 3(b) shows the rotation and tilt angle dependence of $R_P$ values with respect to the optimal values ($10.5°$ for rotation and $2.6°$ for tilt). The error bars for the two angles are estimated by the black dashed line in Fig. 3(b) which is $(R_P)_{\text{error}} = (R_P^{Total})_{\min} + \sigma = 0.35$ [38]. The crossings between the black dashed line and the two curves determine maximum allowed deviations from the optimum solution, yielding $(10.5 \pm 3.0)°$ for rotation angle and $(2.6 \pm 0.8)°$ for tilt angle. These results are consistent with our previous work [33] ($(12\pm3)^\circ$ for rotation and $(2.5\pm1.7)^\circ$ for tilt). The smaller uncertainty in tilt results from using the fractional beams only.

To facilitate the LEED *I-V* structural determination a few assumptions about the structure were made. First, we assumed that octahedra at the surface were not deformed, that means the $\theta_1=\theta_2$ in Fig. 1(d). Second, we assumed that the second layer rotation has the same magnitude and direction as the first layer [26,39]. LEED is not sensitive to activity beyond the first layer, so it is impossible to determine either the direction or magnitude of the rotational distortion in the second layer. In contrast, distortion of octahedra is observed in bulk structural determinations [26,39] and surely will be present at the surface. If we start with the optimal structure and release our constraints on the surface octahedra we find that with a distorted octahedra $R_P$ is slightly reduced and $\theta_1 \neq \theta_2$. We will discuss this in more detail after examining the tilt induced broken symmetry on the LEED spots.

As mentioned earlier the tilt induced broken symmetry makes the intensity of the fractional ordered LEED spots different. For example, the (1/2,5/2) (purple) spot is brighter than the (5/2,1/2) (orange) spot in Fig. 2(b). Fig. 2(c) shows the experimental *I-V* for these beams as a function of incident electron energy at 86 K. The difference between these two beams is $(R_P)_{\mathrm{diff}}=0.68$. The sensitivity of LEED intensity difference as a function of tilt has been explored by conducting a theoretical simulation. *I-V* curves are simulated for the lattice with 10.5° octahedral rotation and various tilt angles. By comparing the simulated *I-V* curves from the purple to the orange fractional spots, the $(R_P)_{\mathrm{diff}}$ is calculated. The simulated $(R_P)_{\mathrm{diff}}$ as a function of tilt angle is shown in the inset of Fig. 2(c). The experimental difference of 0.68 translates into a tilt angle of 4.4°, shown by the red-cross, which is appreciably larger than the value of 2.6° determined by the LEED *I-V* calculation. This is because of our first assumption that the octahedron is rigid, i.e. $\theta_1=\theta_2$. It underestimates the change in $(R_P)_{\mathrm{diff}}$ as the tilt angle increases because it does not allow for the deformation of the octahedra.

As shown in Fig. 1(d), the tilt distortion is described by two angles (Fig. 1(d), the angle between the O3 plane and the *ab*-plane ($\theta_1$) is not necessarily identical to the one between the Ru-O2 bond and the c-axis ($\theta_2$). This unrestricted relaxation gives a slight improvement in $(R_P^{Total})_{\min}$ of 0.01, with $\theta_1 = 2.7°$ and $\theta_2 = 2.3°$. Because of the large error bars, difference between the two angles is not significant.

As the Mn concentration increases the tilt angle may vary. The change in tilt can be quantitatively determined by measuring the intensity of fractional spots such as (3/2,3/2) because their appearance is due to tilt. The LEED pattern of undoped $Sr_3Ru_2O_7$ is shown in Fig. 2(b), and re-plotted in Fig. 4(a). The patterns for the same energy and temperature conditions of *x*=0.01, 0.06 and 0.16 samples are shown in Figs. 4(b)-(d). The line-profile data obtained by tracking the intensity along the white dashed lines in Figs. 4(a)-(d) is shown in Fig. 2(d) for the four different doping levels. All peaks are fit by Gaussian functions, presented by the red and blue shaded regions. The intensity of the fractional spot (3/2,3/2) is normalized based on the averaged intensity of the two integer spots [(1,2) and (2,1)], and the estimated tilt angle is displayed as an inset in Fig. 2(d). The tilt decreases with increasing Mn concentration, and to first order this intensity should be proportional to the tilt angle.

The dynamical behavior of the surface is probed using HREELS. The spectra for *x*=0 and *x*=0.16 samples at 86 K are shown in Fig. 5(a) for specular scattering (for dipole-active excitations). Three phonon modes are observed at $\omega_1$ ~ 30 meV, $\omega_2$ ~ 50 meV, and $\omega_3$ ~ 69.2 meV, consistent with previous measurements on $Sr_2RuO_4$ [40]. These modes can be identified as external cage ($\omega_1$), bending ($\omega_2$), and stretching modes ($\omega_3$). By comparing to the bulk $Sr_3Ru_2O_7$ [41] and surface $Sr_2RuO_4$ [40] measurements, the $\omega_3$ phonon is assigned to the $A_{1g}$ stretching mode of apical oxygen atoms. The movement of atoms in this mode is presented in the inset of Fig. 5(a). For comparison, the corresponding bulk phonon has the energy of 71.5 meV [41], which is slightly higher than the 69.2 meV surface phonon measured by HREELS. Note that in the bulk this is a Raman mode which has symmetric movements of the two apical oxygen atoms. At the surface, the dipole moment emerges from the environment of broken inversion symmetry.

Here we focus on the Mn doping dependence of $\omega_3$ mode. The $\omega_3$ phonon mode at different Mn doping levels is shown in Fig. 5(b). It is conspicuous that the line-shape changes, especially for $x$=0.16 after removing the elastic peak background and other two phonon peaks. This asymmetric line-shape is typical for a Fano process, resulting from the interference between a discrete resonance scattering process ($\omega_3$) and a continuum of background [42]. This Fano line-shape is seen in previous Raman spectroscopy study on $Ca_3Ru_2O_7$ [41,43,44]. In this case, the "resonance" is the phonon excitation, and the "continuum" is the electron-hole pairing excitations. Therefore larger density of states (DOS) near the Fermi level has more excitations, associated with the electron-phonon coupling mechanism which produces an asymmetric line-shape.

The Fano line-shape can be fit with the equation $I(\omega)=I_0(\varepsilon+q)^2/(1+\varepsilon^2)$, where $\varepsilon=(\omega-\omega_0)/\Gamma$, $\omega_0$ is the bare phonon energy, and $\Gamma$ is the intrinsic linewidth. The parameter $q$ is the Fano parameter which measures the line-profile's asymmetry. The smaller the $q$, the more asymmetric the peak line-shape is. The values of $q$ at different doping levels are shown in Fig. 5(b). With increasing Mn concentration the peak becomes more asymmetric, corresponding to smaller $q$. This indicates the surface has higher DOS with larger $x$, i.e., more metallic with less tilt (for higher Mn concentration).

The surface structure-property relationship is summarized in Fig. 6(a), and compared to the bulk in Fig. 6(b), with the color bar indicating the degree of metallicity. In the bulk the physical properties are tied to the octahedral rotational angle. The rotational angle as a function of $x$ looks like what one would expect for a three-dimensional second order phase transition. But for the surface both the tilt angle and Fano $q$ (the measure of metallicity) display linear behavior as a function of $x$ (Fig. 6(a)). This linear relationship mimics the quasi two-dimensional phase transition reported for surface magnetism [45,46].

The surface-induced tilt decreases but rotation maintains large and constant with increasing Mn doping, while in the bulk there is no tilt and the rotation decreases with increasing Mn doping. With such structural difference between the surface and bulk, the physical properties are almost opposite in response to Mn doping. The bulk becomes more insulating when the doping level increases (rotation is removed), but the surface shows an opposite tendency, i.e. becoming more metallic. The tilt distortion at the surface of $Sr_3(Ru_{1-x}Mn_x)_2O_7$ forces the surface into a less metallic state, in the same way as single layer $Ca_{2-x}Sr_xRuO_4$ [16]. The removal of tilt on the surface of $x$=0.16 results in good metallicity similar to that seen in the bulk $Sr_3Ru_2O_7$ ($x$=0): both have octahedral rotation only.

The $Sr_3(Ru_{1-x}Mn_x)_2O_7$ surface enhances rotation and further induces tilt, which are more dramatic than that seen at the surface of $Ca_{2-x}Sr_xRuO_4$. There are two possible reasons. First is the number of layers ($n$) difference between the two compounds. The undoped $n$=1 compound $Sr_2RuO_4$ does not have any rotation or tilt in the bulk, while the undoped $n$=2 compound $Sr_3Ru_2O_7$ already has an octahedral rotational distortion. This indicates that structural instability is enhanced as $n$ increases, so there is more significant distortion for $Sr_4Ru_3O_{10}$ ($n$=3) where the octahedral rotation is large enough to induce ferromagnetism without hydrostatic pressure or magnetic field in the bulk [47]. The octahedra in the outer two layers in the $n$=3 compound are rotated by 5.25° with the central plane having a larger rotation (10.6°) [47]. It is easy to imagine that the presence of the surface could enhance the rotation and may induce a small tilt. The $SrRuO_3$ ($n=\infty$) compound, on the other hand, has intrinsic rotation and tilt up to 820 K and band ferromagnetism [48-51]. As $n$ becomes larger the overall structure is more distorted. The second reason is undoubtedly related to the different doping sites. For $Ca_{2-x}Sr_xRuO_4$ the Ca cation is doped at the alkaline earth metal Sr site (A-site) and it is an isovalent doping, but in $Sr_3(Ru_{1-x}Mn_x)_2O_7$ the Mn cation replaces the transition metal Ru site (B-site) and a Mn atom has one less valence electron than a Ru atom with smaller atomic size. In general in the RP series $A_{n+1}B_nO_{3n+1}$ it is considered that A-site doping affects the structure due to different ionic size, especially on the cage surrounding the octahedra and the c-axis parameter. In contrast, B-site doping affects the $BO_6$ octahedra directly, and the overall structure changes more dramatically. The smaller ionic size of Mn causes the lattice distortion directly associated with the octahedral rotation and tilt. Previous study on $Sr_2RuO_4$ by replacing Ru atoms with 3d transition metal atoms shows that the electronic, magnetic, and structural properties change drastically with even minimum amount of doping [52-55].

The fact that the surface of $Sr_3Ru_2O_7$ is less metallic than the bulk is not surprising. It has been known for almost half a century that creating a surface in any simple three-dimensional material will drive a band narrowing at the surface [45,56,57]. On the other hand, the surface of a semiconductor can be metallic due to the presentence of a surface state. One example known for decades is the metallic 7×7 reconstructed Si(111) surface [58-60]. In contrast, the origin of the structure/property relationship for the surface of $Sr_3(Ru_{1-x}Mn_x)_2O_7$ is quite different than these classical systems: it is the increased rotation angle and induced tilt at the surface that decreases the band width and drive the surface of the parent compound to a more correlated state. The surface rotation for $Sr_3(Ru_{0.84}Mn_{0.16})_2O_7$ will create a new symmetry and a new two-dimensional band structure, like rotation did for $Sr_3Ru_2O_7$ [25].

In summary, we have used the combination of LEED *I-V* and HREELS analysis to investigate the surface phase of $Sr_3(Ru_{1-x}Mn_x)_2O_7$. There is octahedral tilt at the surface for $0 \le x < 0.16$ which does not exist in the bulk, and HREELS measurements indicate that the surface metallicity is strongly coupled to the tilt. The system becomes more conducting with decreasing tilt and enhanced rotation compared to the bulk. The behavior of the surface order parameter versus doping is more or less linear, resembling a second order phase transition at the two-dimensional surface, which is fundamentally different from the three-dimensional like behavior in the bulk.

We thank Wei Chen, Pablo Rivero, and William Shelton for very useful discussions. This work is supported by National Science Foundation Grant DMR-1504226.

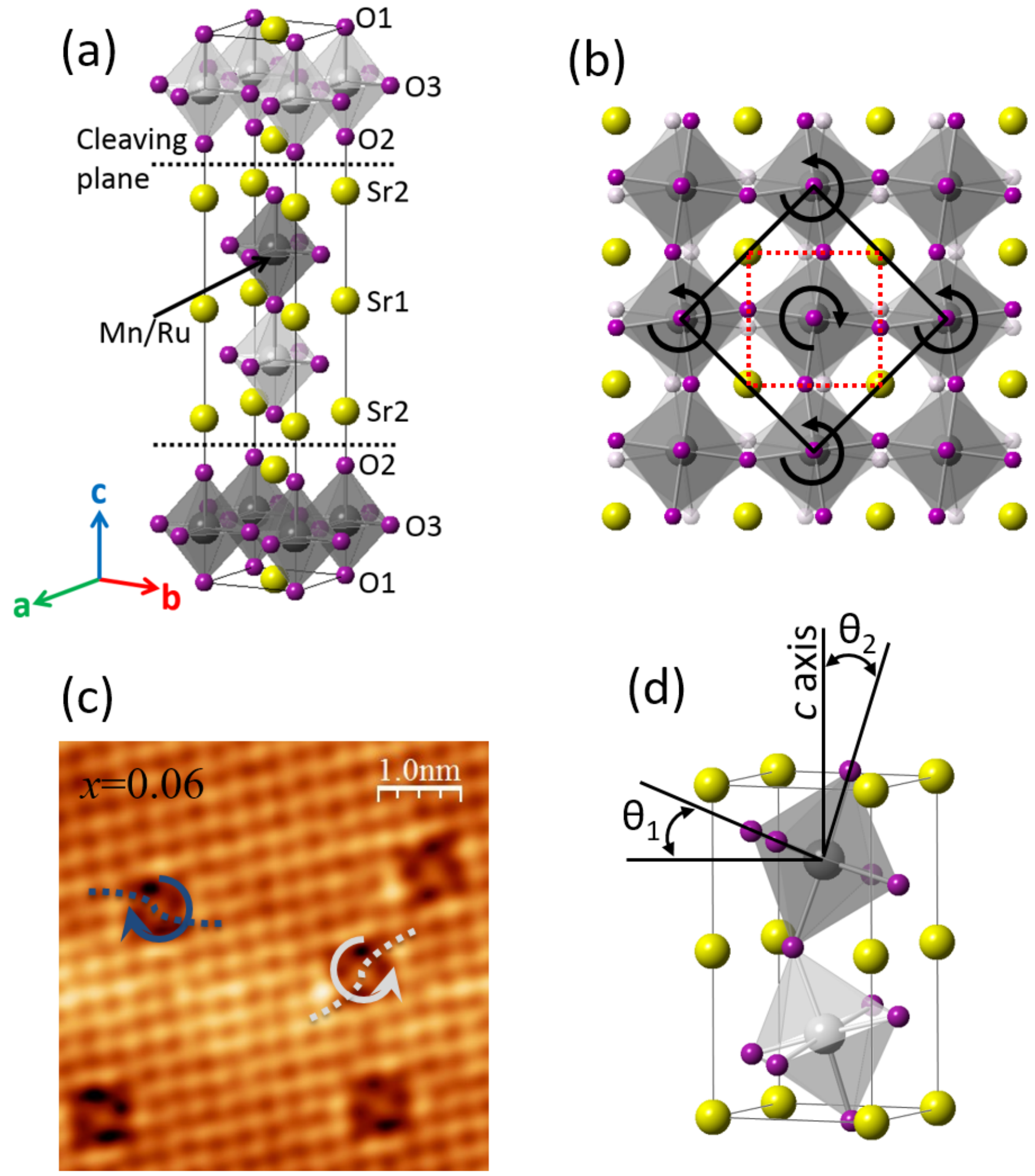

FIG. 1 (color online). (a) Tetragonal unit-cell of $Sr_3(Ru_{1-x}Mn_x)_2O_7$. The cleaving plane is indicated by the dashed line. (b) Planar view of the cleaved surface with octahedral rotation, showing the $(\sqrt{2}\times\sqrt{2})R45°$ unit cell (black), and the tet-$(1\times1)$ unit cell (red). (c) STM topographic image of $x$=0.06 compound (1.2 V bias and 100 K). The different chirality of the Mn sites is indicated by blue and white arrows [34]. (d) Schematic view of octahedral tilt with the tilt angle ($\theta$) defined by either the in-plane angle between the O3 plane and *ab*-plane ($\theta_1$), or the *c*-axis angle between the Ru-O2 bond and *c*-axis ($\theta_2$).

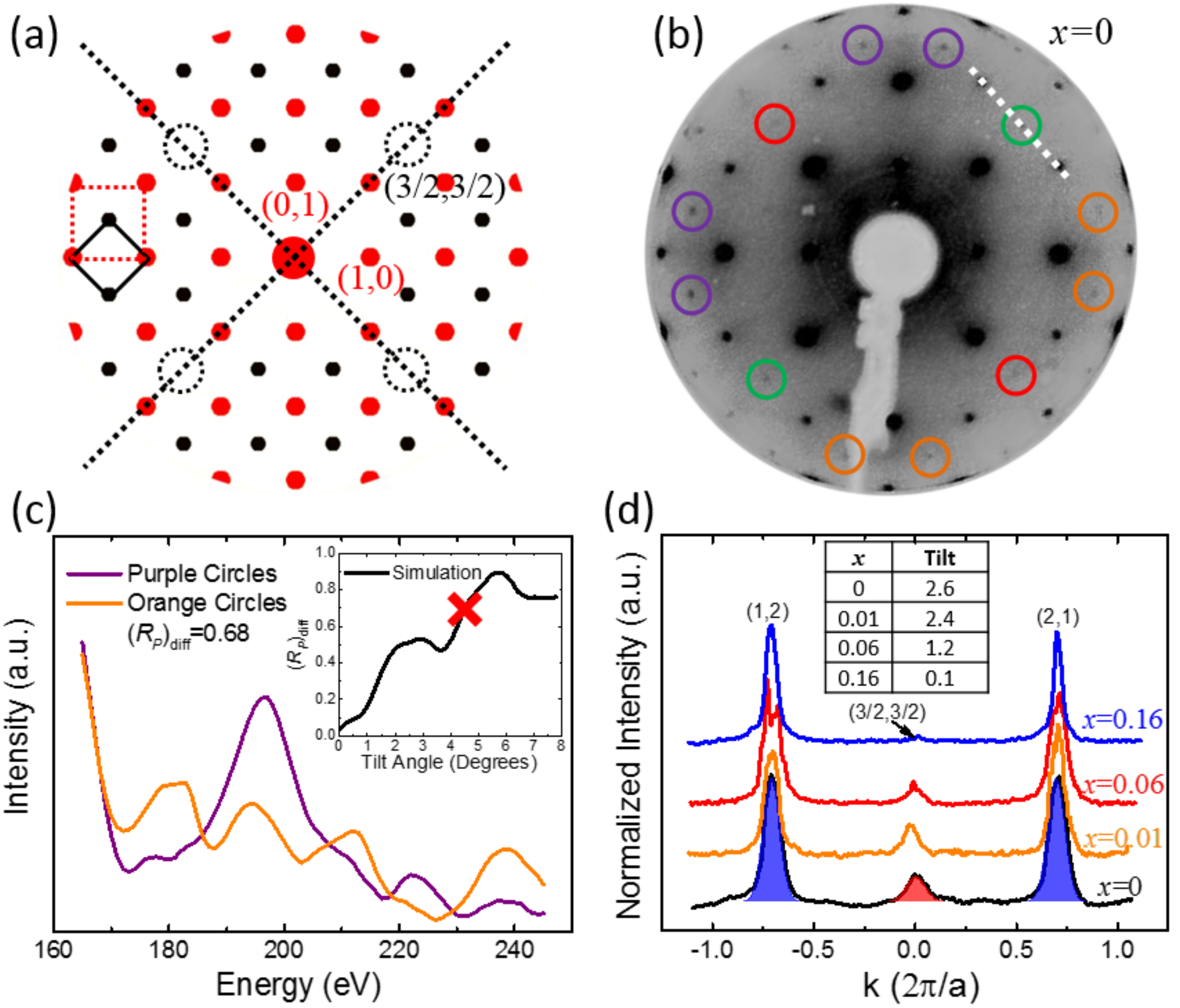

FIG. 2 (color online). (a) Schematic LEED pattern for a surface with octahedral rotation only. The reciprocal lattices from the tet-$(1\times1)$ and $\left(\sqrt{2}\times\sqrt{2}\right)\mathrm{R}45^\circ$ unit cells are indicated by the red dashed square and black solid square, respectively. The two dashed lines are the glidelines where the missing fractional spots (black dashed circles) locate. (b) LEED pattern of $Sr_3Ru_2O_7$ at 86 K and 225 eV. The different colored circles show the spots with different intensities. (c) Comparison of averaged LEED *I*-*V* curves from the purple and orange spots in (b). (Inset) $(R_P)_{\mathrm{diff}}$ values from simulated *I*-*V*'s with various tilt angles, based on 10.5° octahedral rotation. The red cross shows the $(R_P)_{\mathrm{diff}}=0.68$ calculated from experimental *I*-*V* curves. (d) Normalized line profiles along the dashed lines in Fig. 4. The red and blue shaded areas are the intensities of Gaussian fittings, which are used to estimate the tilt angles, shown by the inset table.

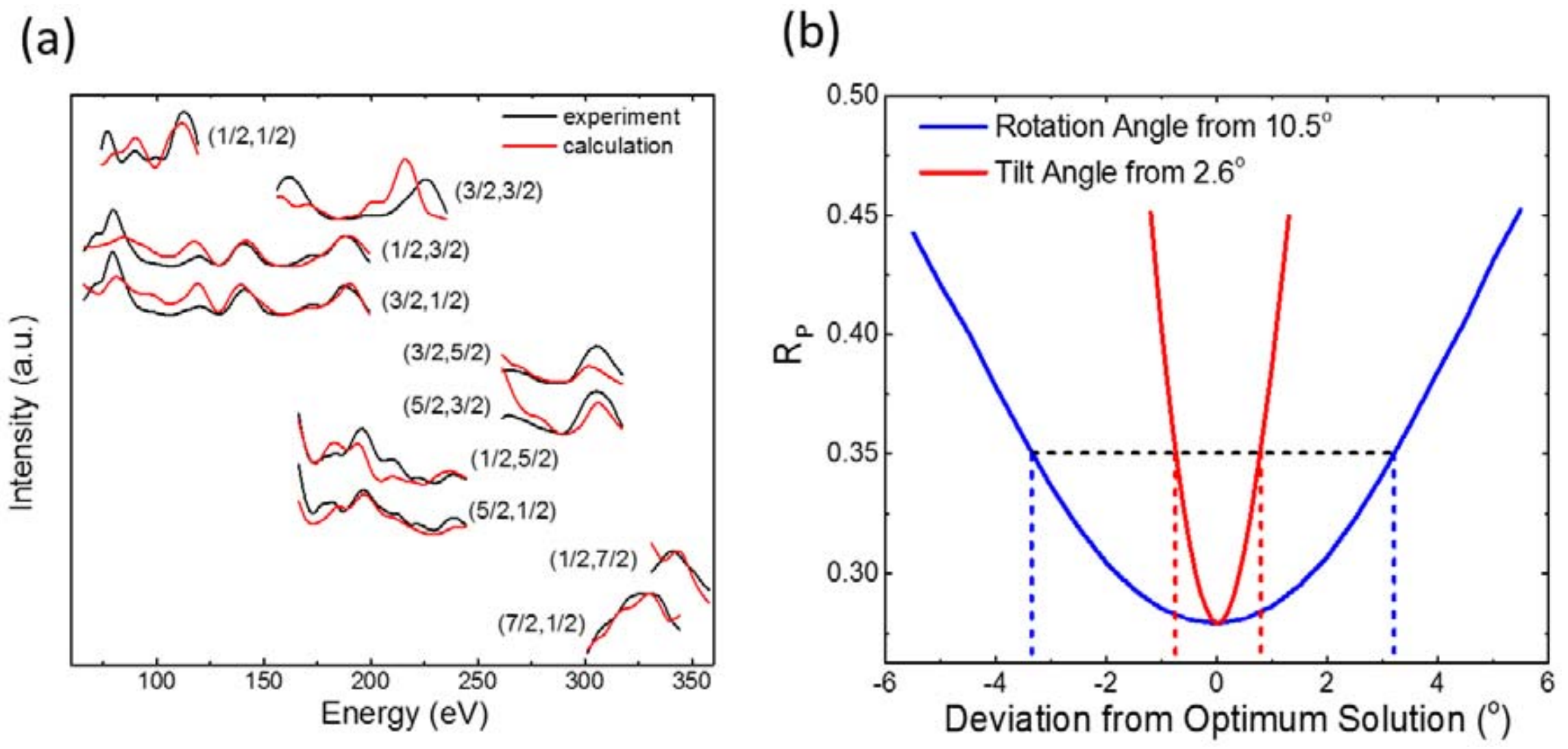

FIG. 3 (color online). (a) Experimental data (black) and calculated LEED I-V curves (red) for fractional spots with various energy ranges. The calculated curves are obtained for the optimized structure with global minimum $R_P$ factor. (b) The $R_P$ factor at angles deviating from the optimum solution, blue for rotation and red for tilt, respectively. The dashed lines are used to estimate the error bars.

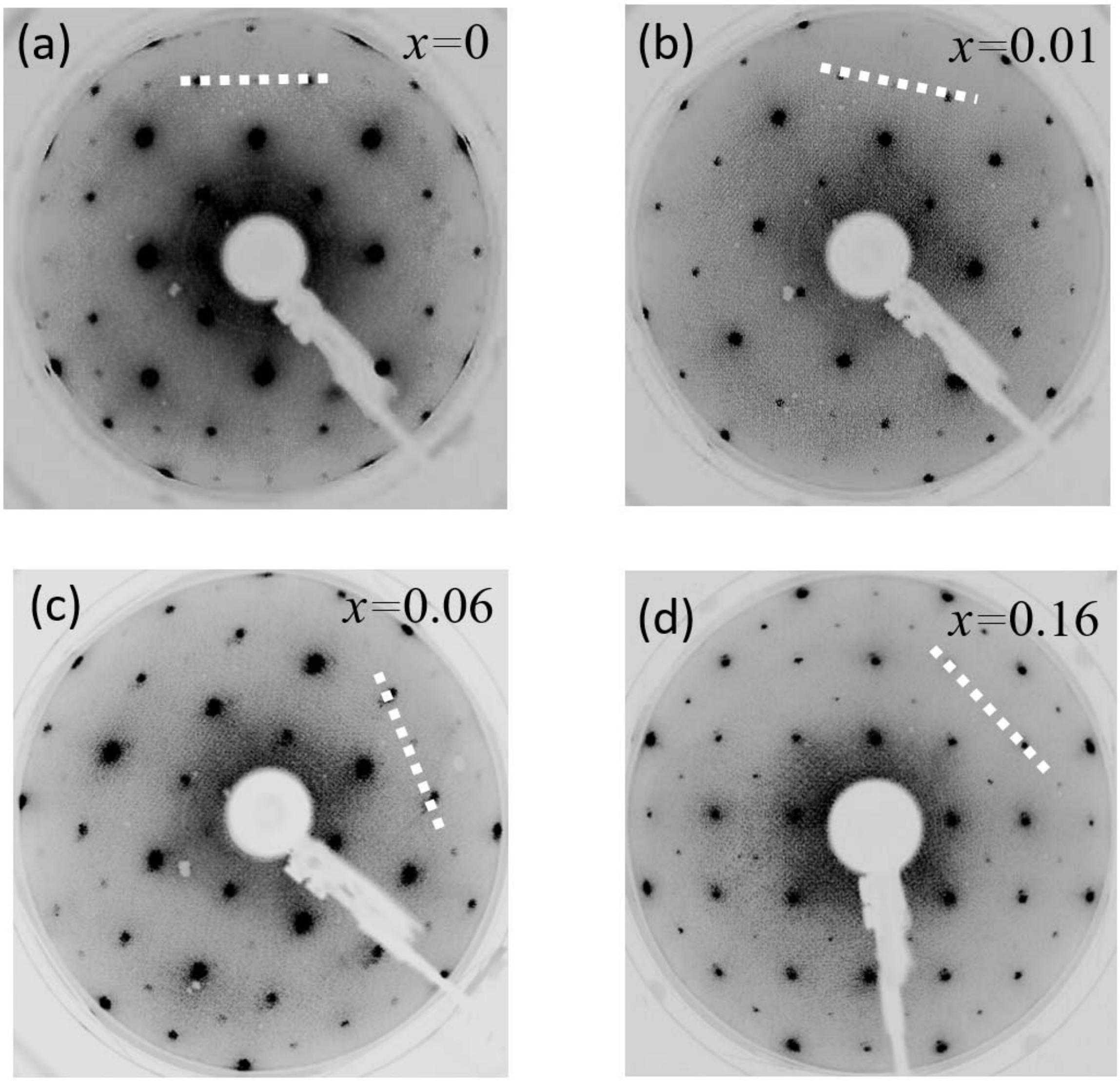

FIG. 4 (color online). LEED patterns taken at the same temperature and beam energy (86 K and 225 eV) for four Mn doped samples with doping levels of $x$=0.00 (a), $x$=0.01 (b), 0.06 (c), 0.16 (d), respectively.

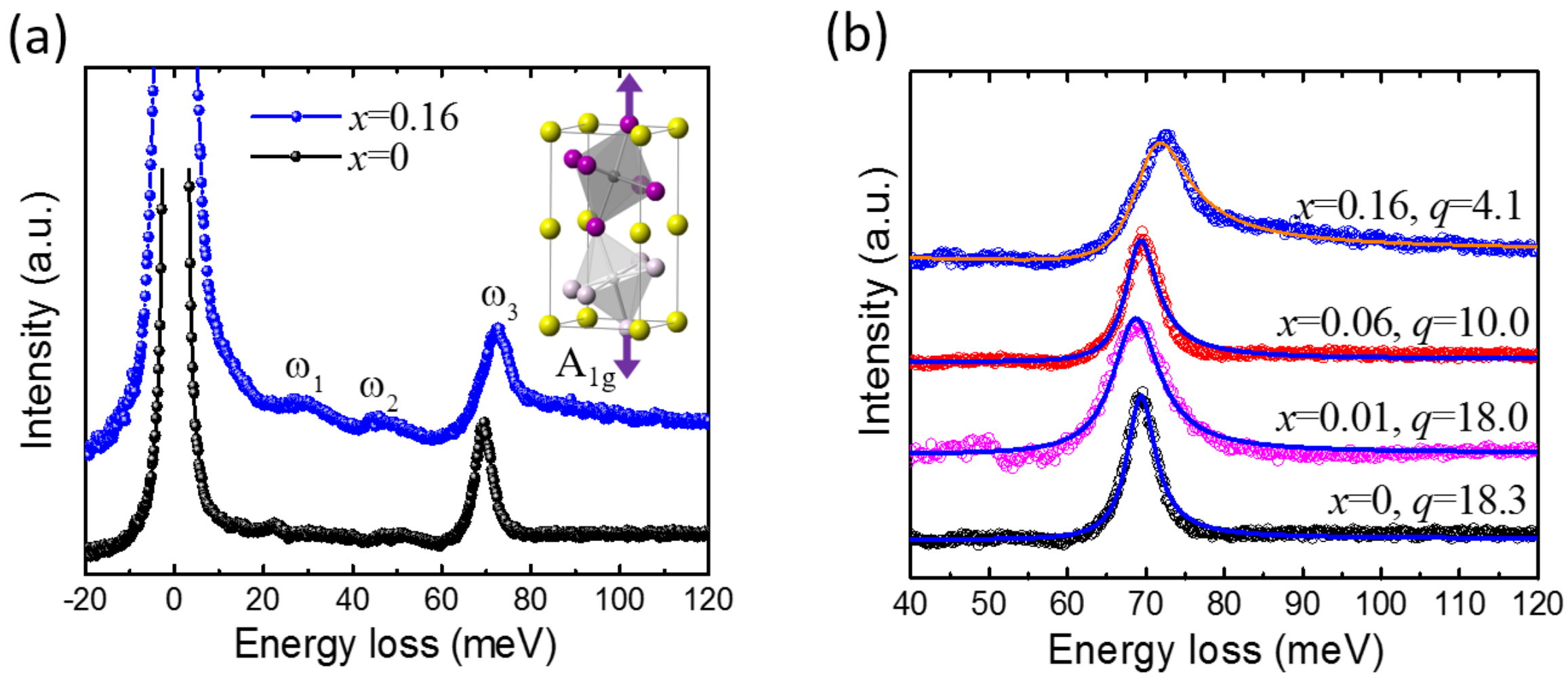

FIG. 5 (color online). (a) HREELS phonon spectra for the surface of $x$=0 and $x$=0.16 samples at 86 K. Inset shows the atomic displacements of the optical phonon mode associated with $\omega_3$. It is an $A_{1g}$ stretching mode of the apical oxygen atoms. (b) The 86 K HREELS data for four different doping levels, with its Fano lineshape fitting after removing the background and other phonons. The line profile becomes more asymmetric when the doping level increases.

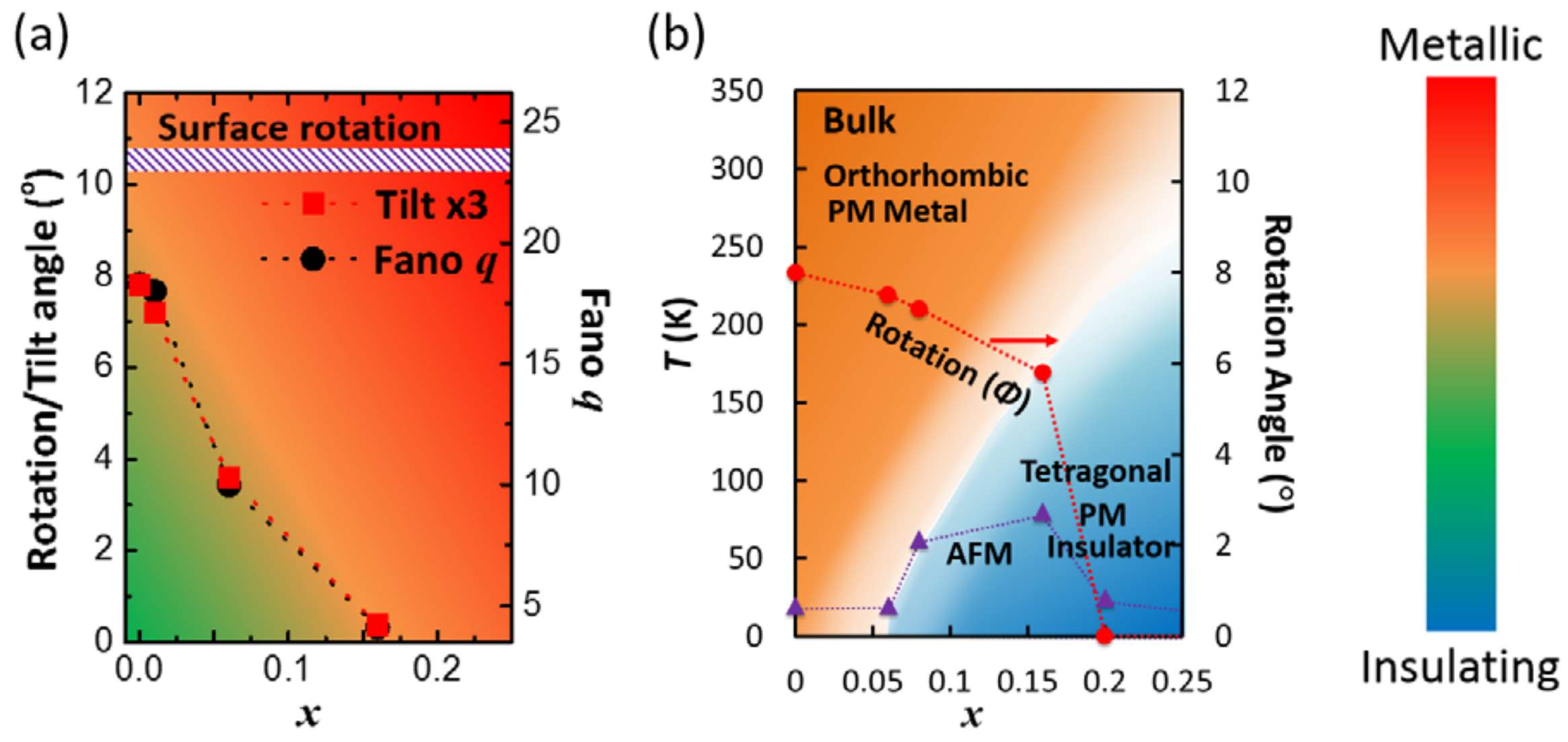

FIG. 6 (color online). (a) Surface phase diagram, comparing the Fano parameter $q$, the tilt angle, and the rotational angle as a function of Mn doping $x$. (b) A portion of the bulk phase diagram. There is a metal-to-insulator transition during the cooling process. The insulating phase becomes more dominant as the doping percentage increases and the octahedral rotation decreases at the same time. The color bar on the right indicates the different degree of metallicity. The surface has an opposite trend in the metallicity compared to the bulk.